\documentclass[twocolumn,prb,showpacs,byrevtex,superbib]{revtex4}
\usepackage{amsmath}
\usepackage{graphicx}
\usepackage{epsfig}
\usepackage[bf]{subfigure}
\usepackage[scaled=.92]{helvet}
\usepackage{courier}
\usepackage{theorem}

\newtheorem{theorem}{Theorem}

{\theorembodyfont{\upshape}
\newtheorem{problem}[theorem]{Problem}
}

{\theorembodyfont{\upshape}

}

\input{tcilatex}
 \DeclareMathAlphabet{\mathsf}{OT1}{cmss}{bx}{n}
\begin{document}

%TCIMACRO{
%\TeXButton{title}{\title
%[Linear Ocean Waves]{A Didactic Approach to Linear Waves in the Ocean}}}%
%BeginExpansion
\title
[Linear Ocean Waves]{A Didactic Approach to Linear Waves in the Ocean}%
%EndExpansion

%TCIMACRO{
%\TeXButton{author}{\author{F. J. Beron-Vera$^{\textrm{a})}$\footnotetext{$^{\textrm{a})}$Electronic mail: \texttt{fberon@rsmas.miami.edu}}}}}%
%BeginExpansion
\author{F. J. Beron-Vera$^{\textrm{a})}$\footnotetext{$^{\textrm{a})}$Electronic mail: \texttt{fberon@rsmas.miami.edu}}}%
%EndExpansion

%TCIMACRO{
%\TeXButton{affiliation}{\affiliation{RSMAS/AMP, University of Miami, Miami, FL 33149}}}%
%BeginExpansion
\affiliation{RSMAS/AMP, University of Miami, Miami, FL 33149}%
%EndExpansion

%TCIMACRO{
%\TeXButton{keywords}{\keywords{Ocean, waves, incompressibility, Boussinesq, hydrostatics, quasigeostrophy.}}}%
%BeginExpansion
%EndExpansion

%TCIMACRO{\TeXButton{pacs}{\pacs{43.30.Bp, 43.30.Cq, 43.30.Ft}}}%
%BeginExpansion
%EndExpansion

%TCIMACRO{\TeXButton{Begin abstract}{\begin{abstract}}}%
%BeginExpansion
\begin{abstract}%
%EndExpansion

The general equations of motion for ocean dynamics are presented and the
waves supported by the (inviscid, unforced) linearized system with respect
to a state of rest are derived. The linearized dynamics sustains one zero
frequency mode (called buoyancy mode) in which salinity and temperature
rearrange in such a way that seawater density does not change. Five nonzero
frequency modes (two acoustic modes, two inertia--gravity or Poincar\'{e}
modes, and one planetary or Rossby mode) are also sustained by the
linearized dynamics, which satisfy an asymptotic general dispersion
relation. The most usual approximations made in physical oceanography
(namely incompressibility, Boussinesq, hydrostatic, and quasigeostrophic)
are also consider, and their implications in the reduction of degrees of
freedom (number of independent dynamical fields or prognostic equations) of,
and compatible waves with, the linearized governing equations are
particularly discussed and emphasized.

%TCIMACRO{\TeXButton{End abstract}{\end{abstract}}}%
%BeginExpansion
\end{abstract}%
%EndExpansion

%TCIMACRO{\TeXButton{volumeyear}{\volumeyear{2002}}}%
%BeginExpansion
\volumeyear{2002}%
%EndExpansion

%TCIMACRO{\TeXButton{volumenumber}{\volumenumber{205}}}%
%BeginExpansion
\volumenumber{205}%
%EndExpansion

%TCIMACRO{\TeXButton{issuenumber}{\issuenumber{5}}}%
%BeginExpansion
\issuenumber{5}%
%EndExpansion

%TCIMACRO{\TeXButton{date}{\date[Dated: ]{\today}}}%
%BeginExpansion
\date[Dated: ]{\today}%
%EndExpansion

%TCIMACRO{\TeXButton{startpage}{\startpage{1}}}%
%BeginExpansion
\startpage{1}%
%EndExpansion

%TCIMACRO{\TeXButton{endpage}{\endpage{102}}}%
%BeginExpansion
\endpage{102}%
%EndExpansion

%TCIMACRO{\TeXButton{eid}{\eid{identifier}}}%
%BeginExpansion
\eid{identifier}%
%EndExpansion

%TCIMACRO{\TeXButton{maketitle}{\maketitle}}%
%BeginExpansion
\maketitle%
%EndExpansion

\section{Introduction}

The goal of this educational work is to show how the various types of linear
waves in the ocean (acoustic, inertia--gravity or Poincar\'{e}, and
planetary or Rossby waves) can be obtained from a general dispersion
relation in an approximate (asymptotic) sense. Knowledge of the theory of
partial differential equations, and basic classical and fluid mechanics are
only needed for the reader to understand the material presented here, which
could be taught as a special topic in a course of fluid mechanics for
physicists.

The exposition starts by presenting the general equations of motion for
ocean dynamics in Sec. \ref{eqs}. This presentation is not intended to be
rigorous, but rather conceptual. Accordingly, the equations of motion are
simplified as much as possible for didactic purposes. The general dispersion
relation for the waves supported by the (inviscid, unforced) linearized
dynamics with respect to a quiescent state is then derived in Sec. \ref%
{ondas}. This is done by performing a separation of variables between the
vertical direction, on one side, and the horizontal position and time, on
the other side. Sec. \ref{discussion} discusses the implications of the most
common approximations made in oceanography (namely incompressibility,
Boussinesq, hydrostatic, and quasigeostrophic) in the reduction of degrees
of freedom (number of independent dynamical fields or prognostic equations)
of, and compatible waves with, the linearized governing equations.
Particular emphasis is made on this important issue, which is vaguely
covered in standard textbooks (e.g. Refs.
%TCIMACRO{
%\TeXButton{LeBlond-Mysak-78,Gill-82,Pedlosky-87}{\onlinecite{LeBlond-Mysak-78,Gill-82,Pedlosky-87}}}%
%BeginExpansion
\onlinecite{LeBlond-Mysak-78,Gill-82,Pedlosky-87}%
%EndExpansion
). Some problems have been interspersed within the text to help the reader
to assimilate the material presented. The solutions to some of these
problems are outlined in App. \ref{answers}.

\section{General Equations of Motion\label{eqs}}

Let $\mathbf{x}:=(x,y)$ be the horizontal position, i.e. tangential to the
Earth's surface, with $x$ and $y$ its eastward and northward components,
respectively; let $z$ be the upward coordinate; and let $t$ be time. Unless
otherwise stated all variables are functions of $(\mathbf{x,}z,t)$ in this
paper.

The \textit{thermodynamic state} of the ocean is determined by three
variables, typically $S$ (salinity), $T$ (temperature), and $p$ (pressure,
referred to one atmosphere). Seawater density, $\rho ,$ is a function of
these three variables, i.e. $\rho =\rho (S,T,p),$ known as the state
equation of seawater. In particular,%
\begin{equation}
\rho ^{-1}\mathrm{D}\rho =\alpha _{S}\mathrm{D}S-\alpha _{T}\mathrm{D}%
T+\alpha _{p}\mathrm{D}p.  \label{estado}
\end{equation}%
Here, $\mathrm{D}:=\partial _{t}+\mathbf{u}\cdot \mathbf{\nabla }+w\partial
_{z}\mathrm{\ }$is the substantial or material derivative, where $\mathbf{u}$
and $w$ are the horizontal and vertical components of the velocity field,
respectively, and $\mathbf{\nabla }$ denotes the horizontal gradient; $%
\alpha _{S}:=\rho ^{-1}\left( \partial _{S}\rho \right) _{T,p}$ and $\alpha
_{T}:=\rho ^{-1}\left( \partial _{T}\rho \right) _{S,p}$ are the haline
contraction and thermal expansion coefficients, respectively; and $\alpha
_{p}:=\rho ^{-1}\left( \partial _{p}\rho \right) _{T,S}=\alpha _{T}\Gamma
+\rho ^{-1}c_{\mathrm{s}}^{-2},$ where $\Gamma $ is the adiabatic gradient
and $c_{\mathrm{s}}$ is the speed of sound, which characterize the
compressibility of seawater.

The \textit{physical state} of the ocean is determined at every instant by
the above three variables $(S,T,p)$ and the three components of the velocity
field $(\mathbf{u},w),$ i.e. \textit{six }independent scalar\textit{\ }%
variables. The evolution of these variables are controlled by
\begin{subequations}
\label{sys1}
\begin{eqnarray}
\mathrm{D}S\hspace{-0.05in} &=&\hspace{-0.05in}F_{S}, \\
\mathrm{D}T\hspace{-0.05in} &=&\hspace{-0.05in}\Gamma \mathrm{D}p+F_{T}, \\
\mathrm{D}p\hspace{-0.05in} &=&\hspace{-0.05in}-\alpha _{p}^{-1}\left(
\mathbf{\nabla }\cdot \mathbf{u}+\partial _{z}w+\alpha _{T}F_{T}-\alpha
_{S}F_{S}\right) , \\
\mathrm{D}\mathbf{u}\hspace{-0.05in} &=&\hspace{-0.05in}-\rho ^{-1}\mathbf{%
\nabla }p+\mathbf{F}_{\mathbf{u}}, \\
\mathrm{D}w\hspace{-0.05in} &=&\hspace{-0.05in}-\rho ^{-1}\mathbf{\partial }%
_{z}p-g+F_{w}.
\end{eqnarray}%
In Newton's horizontal equation (\ref{sys1}d), the term $\mathbf{F}_{\mathbf{%
u}}$ represents the acceleration due to the horizontal components of the
Coriolis and frictional forces. In Newton's vertical equation (\ref{sys1}d),
the term $F_{w}$ represents the acceleration due to the vertical component
of the Coriolis and frictional forces, and $g$ is the (constant)
acceleration due to gravity. The term $F_{S}$ in the salinity equation (\ref%
{sys1}a) represents diffusive processes. The term $F_{T}$ in the thermal
energy equation (\ref{sys1}b), which follows from the first principle of
thermodynamics, represents the exchange of heat by conduction and radiation,
as well as heating by change of phase, chemical reactions or viscous
dissipation. The pressure or continuity equation (\ref{sys1}c) follows from (%
\ref{estado}).

\begin{problem}
Investigate why (\ref{sys1}d,e) do not include the centrifugal force which
would also be needed to describe the dynamics in a noninertial reference
frame such as one attached to the Earth.
\end{problem}

Since adiabatic compression does not have important dynamical effects, in
physical oceanography it is commonly neglected. This is accomplished upon
introducing the potential temperature $\theta $, which satisfies $\alpha
_{\theta }\mathrm{D}\theta =\alpha _{T}(\mathrm{D}T-\Gamma \mathrm{D}p),$ so
that $\rho =\rho (S,\theta ,p)$ and (\ref{estado}) is consistently replaced
by
\end{subequations}
\begin{equation}
\rho ^{-1}\mathrm{D}\rho =\alpha _{S}\mathrm{D}S-\alpha _{\theta }\mathrm{D}%
\theta +\rho ^{-1}c_{\mathrm{s}}^{-2}\mathrm{D}p,
\end{equation}%
where here it must be understood that $\alpha _{S}=(\partial _{S}\rho
)_{\theta ,p}$ and $\alpha _{T}=(\partial _{\theta }\rho )_{S,p}.$ Equations
(\ref{sys1}b,c) then are replaced, respectively, by
\begin{subequations}
\label{sys2}
\begin{eqnarray}
\mathrm{D}\theta \hspace{-0.05in} &=&\hspace{-0.05in}F_{\theta }, \\
\mathrm{D}p\hspace{-0.05in} &=&\hspace{-0.05in}-\rho c_{\mathrm{s}%
}^{2}\left( \mathbf{\nabla }\cdot \mathbf{u}+\partial _{z}w+\alpha _{\theta
}F_{\theta }-\alpha _{S}F_{S}\right) .
\end{eqnarray}

As our interest is in the waves sustained by the linearized dynamics, we do
not need to consider either diffusive processes or allow the motion to
depart from isentropic. Hence, we will set $F_{S}\equiv 0\equiv F_{\theta }$
so that equations (\ref{sys1}a) and (\ref{sys2}) can be substituted,
respectively, by
\end{subequations}
\begin{subequations}
\label{sys3}
\begin{eqnarray}
\mathrm{D}\zeta \hspace{-0.05in} &=&\hspace{-0.05in}w, \\
\mathrm{D}p\hspace{-0.05in} &=&\hspace{-0.05in}-\rho c_{\mathrm{s}%
}^{2}\left( \mathbf{\nabla }\cdot \mathbf{u}+\partial _{z}w\right) ,
\end{eqnarray}%
where $\zeta $ is the vertical displacement of an isopycnal which is defined
such that $\rho =\rho _{\mathrm{r}}(z-\zeta )$.

\begin{problem}
Show that equations (\ref{sys1}a) and (\ref{sys2}a) certainly lead to (\ref%
{sys3}a) when $F_{S}\equiv 0\equiv F_{\theta }.$
\end{problem}

We will also neglect frictional effects, so that equations (\ref{sys1}d,e)
are seen to be nothing but Euler equations of (ideal) fluid mechanics with
the addition of the Coriolis force. The latter will be further considered as
due solely to the vertical component of the Earth rotation. Thus, the
following simplified form of equations (\ref{sys1}d,e) will be considered:
\end{subequations}
\begin{subequations}
\label{sys4}
\begin{eqnarray}
\mathrm{D}\mathbf{u}\hspace{-0.05in} &=&\hspace{-0.05in}-\rho ^{-1}\mathbf{%
\nabla }p-f\mathbf{\hat{z}}\times \mathbf{u}, \\
\mathrm{D}w\hspace{-0.05in} &=&\hspace{-0.05in}-\rho ^{-1}\mathbf{\partial }%
_{z}p-g.
\end{eqnarray}%
Here, $\mathbf{\hat{z}}$ is the upward unit vector and $f:=2\Omega \sin
\vartheta $, where $\Omega $ is the (assumed constant) spinning rate of the
Earth around its axis and $\vartheta $ is the geographical latitude, is the
Coriolis parameter. For simplicity, we will avoid working in full spherical
geometry. Instead, we will consider $f=f_{0}+\beta y,$ where $f_{0}:=2\Omega
\sin \vartheta _{0}$ and $\beta :=2\Omega R^{-1}\cos \vartheta _{0}$ with $%
\vartheta _{0}$ a fixed latitude and $R$ the mean radius of the planet, and $%
\mathbf{\nabla }=(\partial _{x},\partial _{y}),$ which is known as the $%
\beta $-plane approximation. It should remain clear, however, that a
consistent $\beta $-plane approximation must include some geometric
(non-Cartesian) terms. \cite{Ripa-JPO-97b} Neither these terms nor those of
the Coriolis force due to the horizontal component of the Earth's rotation
contribute to add waves to the linearized equations of motion. Their
neglection is thus well justified for the purposes of this paper.

\section{Waves of the Linearized Dynamics\label{ondas}}

Consider a state of rest ($\mathbf{u}=\mathbf{0},$ $w=0$) characterized by $%
\mathrm{d}p_{\mathrm{r}}/\mathrm{d}z=-\rho _{\mathrm{r}}g$, where $p_{%
\mathrm{r}}(z)$ and $\rho _{\mathrm{r}}(z)$ are reference profiles of
pressure and density, respectively. Equations (\ref{sys2}) and (\ref{sys4}),
linearized with respect to that state, can be written as
\end{subequations}
\begin{subequations}
\label{sis1}
\begin{eqnarray}
\partial _{t}\zeta ^{\prime }\hspace{-0.05in} &=&\hspace{-0.05in}w^{\prime },
\\
\partial _{t}p^{\prime }\hspace{-0.05in} &=&\hspace{-0.05in}-\rho _{\mathrm{r%
}}c_{\mathrm{s}}^{2}\left( \mathbf{\nabla }\cdot \mathbf{u}^{\prime
}+\partial _{z}^{-}w^{\prime }\right) , \\
\partial _{t}\mathbf{u}^{\prime }\hspace{-0.05in} &=&\hspace{-0.05in}-\rho _{%
\mathrm{r}}^{-1}\mathbf{\nabla }p^{\prime }-f\mathbf{\hat{z}}\times \mathbf{u%
}^{\prime }, \\
\partial _{t}w^{\prime }\hspace{-0.05in} &=&\hspace{-0.05in}-\rho _{\mathrm{r%
}}^{-1}\partial _{z}^{+}p^{\prime }-N^{2}\zeta ^{\prime }.
\end{eqnarray}%
Here, primed quantities denote perturbations with respect to the state of
rest; $\partial _{z}^{\pm }:=\partial _{z}\pm gc_{\mathrm{s}}^{-2}$; $%
N^{2}(z):=-g(\rho _{\mathrm{r}}^{-1}\mathrm{d}\rho _{\mathrm{r}}/\mathrm{d}%
z+gc_{\mathrm{s}}^{-2})$ is the square of the reference Brunt-V\"{a}is\"{a}l%
\"{a} frequency; and $c_{\mathrm{s}}$ is assumed constant. In addition to
the above equations, it is clear that
\end{subequations}
\begin{subequations}
\label{ST}
\begin{eqnarray}
\partial _{t}S^{\prime }\hspace{-0.05in} &=&\hspace{-0.05in}-w\,\mathrm{d}S_{%
\mathrm{r}}/\mathrm{d}z, \\
\partial _{t}\theta ^{\prime }\hspace{-0.05in} &=&\hspace{-0.05in}-w\,%
\mathrm{d}\theta _{\mathrm{r}}/\mathrm{d}z,
\end{eqnarray}%
where $S_{\mathrm{r}}(z)$ and $\theta _{\mathrm{r}}(z)$ are reference
salinity and potential density profiles, respectively.

\begin{problem}
Work out the linearization of the equations of motion.
\end{problem}

\subsection{Zero Frequency Mode}

The linearized dynamics supports a solution with vanishing frequency ($%
\partial _{t}\equiv 0$) such that
\end{subequations}
\begin{equation}
\zeta ^{\prime }\equiv 0,\quad p^{\prime }\equiv 0,\quad \mathbf{u}^{\prime
}\equiv \mathbf{0},\quad w^{\prime }\equiv 0,  \label{w=0}
\end{equation}%
as it follows from (\ref{sis1}), but with
\begin{equation}
S^{\prime }\neq 0,\quad \theta ^{\prime }\neq 0,
\end{equation}%
as can be inferred from (\ref{ST}). Namely for this solution the salinity
and temperature fields vary without changing the density of the fluid. More
precisely, one has, on one hand, $\rho ^{\prime }=\rho _{\mathrm{r}%
}(g^{-1}N^{2}+gc_{\mathrm{s}}^{-2})\zeta ^{\prime }$, and, on the other, $%
\rho ^{\prime }=\rho _{\mathrm{r}}(\alpha _{S}S^{\prime }-\alpha _{\theta
}\theta ^{\prime }+\alpha _{p}p^{\prime }),$ where the $\alpha $'s are
evaluated at the reference state. By virtue of (\ref{w=0}) then it follows
that
\begin{equation}
\alpha _{S}S^{\prime }-\alpha _{\theta }\theta ^{\prime }=0.
\end{equation}%
This so-called \textit{buoyancy mode} describes small scale processes in the
ocean such as double diffusion.

\subsection{Nonzero Frequency Modes}

Upon eliminating $w$ between (\ref{sis1}a) and (\ref{sis1}d), and proposing
a separation of variables between $z$, on one side, and $(\mathbf{x},t),$ on
the other side, for the horizontal velocity and pressure fields in the form%
\begin{equation}
\mathbf{u}^{\prime }=\mathbf{u}^{c}(\mathbf{x,}t)\,\partial _{z}^{-}F(z),%
\text{\quad }p^{\prime }=\rho _{\mathrm{r}}(z)\,p^{c}(\mathbf{x}%
,t)\,\partial _{z}^{-}F(z),
\end{equation}%
it follows that
\begin{subequations}
\label{sis2}
\begin{eqnarray}
c_{\mathrm{s}}^{-2}\partial _{z}^{-}F\partial _{t}p^{c}+\left( \partial
_{z}^{-}F\mathbf{\nabla }\cdot \mathbf{u}^{c}+\partial _{zt}^{-}\zeta
^{\prime }\right) \hspace{-0.05in} &=&\hspace{-0.05in}0, \\
\partial _{t}\mathbf{u}^{c}+f\mathbf{\hat{z}}\times \mathbf{u}^{c}+\mathbf{%
\nabla }p^{c}\hspace{-0.05in} &=&\hspace{-0.05in}\mathbf{0}, \\
\partial _{tt}\zeta ^{\prime }+\rho _{\mathrm{r}}^{-1}\partial
_{z}^{+}\left( \rho _{\mathrm{r}}\partial _{z}^{-}F\right) p^{c}+N^{2}\zeta
^{\prime }\hspace{-0.05in} &=&\hspace{-0.05in}0.
\end{eqnarray}%
Now, assuming a common temporal dependence of the form $\mathrm{e}^{-\mathrm{%
i}\omega t}$, from (\ref{sis2}c) one obtains
\end{subequations}
\begin{equation}
\zeta ^{\prime }=-\frac{p^{c}\partial _{z}^{+}\left( \rho _{\mathrm{r}%
}\partial _{z}^{-}F\right) }{\rho _{\mathrm{r}}\left( N^{2}-\omega
^{2}\right) }.  \label{zeda}
\end{equation}%
Then, upon substituting (\ref{zeda}) in (\ref{sis2}a) it follows that
\begin{equation}
c_{\mathrm{s}}^{-2}-\frac{1}{\partial _{z}^{-}F}\partial _{z}^{-}\left[
\frac{\partial _{z}^{+}\left( \rho _{\mathrm{r}}\partial _{z}^{-}F\right) }{%
\rho _{\mathrm{r}}\left( N^{2}-\omega ^{2}\right) }\right] =-\frac{\mathbf{%
\nabla }\cdot \mathbf{u}^{c}}{\partial _{t}p^{c}}=c^{-2},  \label{sep}
\end{equation}%
where $c$ is a constant known as the \textit{separation constant}. Clearly, $%
c_{\mathrm{s}}$ must be chosen as a constant in order for the separation of
variables to be possible. From (\ref{sep}) it follows, on one hand, that
\begin{equation}
\partial _{z}^{+}\left( \rho _{\mathrm{r}}\partial _{z}^{-}F\right) +\rho _{%
\mathrm{r}}\left( N^{2}-\omega ^{2}\right) \left( c^{-2}-c_{\mathrm{s}%
}^{-2}\right) F=0,  \label{vert}
\end{equation}%
and taking into account (\ref{sis2}b), it follows, on the other hand, that
\begin{subequations}
\label{hor}
\begin{eqnarray}
\partial _{t}p^{c}+c^{2}\mathbf{\nabla }\cdot \mathbf{u}^{c}\hspace{-0.05in}
&=&\hspace{-0.05in}0, \\
\partial _{t}\mathbf{u}^{c}+f\mathbf{\hat{z}}\times \mathbf{u}^{c}+\mathbf{%
\nabla }p^{c}\hspace{-0.05in} &=&\hspace{-0.05in}\mathbf{0}.
\end{eqnarray}%
Equation (\ref{vert}) governs the \textit{vertical structure} of the
perturbations, whereas system (\ref{hor}) controls the evolution of their
\textit{horizontal structure}.

\begin{problem}
Show that the alternate separation of variables which uses $F(z)$ instead of
$\partial _{z}^{-}F(z)$ leads to a vertical structure equation with a
singularity where $\omega ^{2}\equiv N^{2}.$
\end{problem}

Equation (\ref{vert}) can be presented in different forms according to the
approximation performed. Under the \textit{incompressibility approximation},
which consists of making the replacement $\partial _{t}p^{\prime
}+gw^{\prime }\mapsto 0$ in the continuity equation (\ref{sis1}b), equation (%
\ref{vert}) takes the form
\end{subequations}
\begin{equation}
\partial _{z}^{+}\left( \rho _{\mathrm{r}}\partial _{z}F\right) +\rho _{%
\mathrm{r}}c^{-2}\left( N^{2}-\omega ^{2}\right) F=0.
\end{equation}%
This approximation corresponds formally to taking the limit $c_{\mathrm{s}%
}^{-2}\rightarrow 0.$ The \textit{hydrostatic approximation}, in turn,
consists of making the replacement $\partial _{t}w^{\prime }\mapsto 0$ in
Newton's vertical equation (\ref{sis1}d). This way, without the need of
assuming any particular temporal dependence, it follows that $\zeta ^{\prime
}=-p^{c}\partial _{z}^{+}\left( \rho _{\mathrm{r}}\partial _{z}^{-}F\right)
/(\rho _{\mathrm{r}}N^{2})$. Consequently, equation (\ref{vert}) reduces to
\begin{equation}
\partial _{z}^{+}\left( \rho _{\mathrm{r}}\partial _{z}^{-}F\right) +\rho _{%
\mathrm{r}}\left( c^{-2}-c_{\mathrm{s}}^{-2}\right) N^{2}F=0.
\end{equation}%
This approximation is valid for $\omega ^{2}\ll N^{2},$ i.e. periods
exceeding the local buoyancy period which typically is of about $1$ \textrm{h%
}. (Of course, this approximation implies that of incompressibility as it
filters out the acoustic modes whose frequencies are much higher than the
Brunt-V\"{a}is\"{a}l\"{a} frequency.) Another common approximation is the
\textit{Boussinesq approximation}, which consists of making the replacements
$\rho _{\mathrm{r}}\mapsto \bar{\rho}=\mathrm{const}$\textrm{.} and $%
\partial _{z}^{\pm }\mapsto \partial _{z}$ in (\ref{sis1}). Under this
approximation, equation takes the simpler form
\begin{equation}
\mathrm{d}^{2}F/\mathrm{d}z^{2}+\left( c^{-2}-c_{\mathrm{s}}^{-2}\right)
\left( N^{2}-\omega ^{2}\right) F=0.  \label{Boussinesq}
\end{equation}

\begin{problem}
Show that the Boussinesq approximation is very good for the ocean but not so
for the atmosphere. Hint: This approximation requires $c_{\mathrm{s}}^{2}\gg
gH,$ where $H$ is a typical vertical length scale.
\end{problem}

\subsubsection{Horizontal Structure}

To describe these waves is convenient to introduce a potential $\varphi (%
\mathbf{x},t)$ such that \cite{Ripa-JFM-94,Ripa-UGM-97}
\begin{subequations}
\begin{eqnarray}
p^{c}\hspace{-0.05in} &=&\hspace{-0.05in}-c^{2}\left( \partial
_{ty}+f\partial _{x}\right) \varphi , \\
u^{c}\hspace{-0.05in} &=&\hspace{-0.05in}\left( c^{2}\partial
_{xy}+f\partial _{t}\right) \varphi , \\
v^{c}\hspace{-0.05in} &=&\hspace{-0.05in}\left( \partial _{tt}-c^{2}\partial
_{xx}\right) \varphi ,
\end{eqnarray}%
which allows one to reduce system (\ref{hor}) to a single equation in one
variable:
\end{subequations}
\begin{equation}
\mathcal{L}\varphi :=\left\{ \partial _{t}\left[ \partial
_{tt}+f^{2}(y)-c^{2}\nabla ^{2}\right] -\beta c^{2}\partial _{x}\right\}
\varphi =0.  \label{Ophi}
\end{equation}%
The linear differential operator $\mathcal{L}$ contains a variable
coefficient and, hence, a solution to (\ref{Ophi}) must be of the form%
\begin{equation}
\varphi =\Phi (y)\mathrm{e}^{\mathrm{i}\left( kx-\omega t\right) }
\end{equation}%
with $\Phi (y)$ satisfying
\begin{equation}
\mathrm{d}^{2}\Phi /\mathrm{d}y^{2}+l^{2}(y)\Phi =0  \label{phi}
\end{equation}%
where
\begin{equation}
l^{2}(y):=-k^{2}-\beta \frac{k}{\omega }+\frac{\omega ^{2}-f^{2}(y)}{c^{2}}.
\label{l}
\end{equation}%
Now, if $l^{2}(y)$ is positive and sufficiently large, then $\Phi (y)$
oscillates like%
\begin{equation}
\Phi (y)\sim \mathrm{e}^{\pm \mathrm{i}\tint^{y}\mathrm{d}y\,l(y)}.
\end{equation}%
This is known as the WKB approximation (cf. e.g. Ref.
%TCIMACRO{\TeXButton{Olver-74}{\onlinecite{Olver-74}}}%
%BeginExpansion
\onlinecite{Olver-74}%
%EndExpansion
), where $l(y)$ defines a local meridional wavenumber in the approximate
(asymptotic) dispersion relation
\begin{equation}
\omega ^{2}-(f^{2}+c^{2}\mathbf{k}^{2})-\beta \frac{k}{\omega }=0,
\label{w(k,l)}
\end{equation}%
where $\mathbf{k}:=(k,l)$ is the horizontal wavenumber.

System (\ref{hor}) also supports a type of nondispersive waves called Kelvin
waves. These waves have $v^{c}\equiv 0$ and thus are seen to satisfy
\begin{subequations}
\label{v=0}
\begin{eqnarray}
\partial _{t}p^{c}+c^{2}\partial _{x}u^{c}\hspace{-0.05in} &=&\hspace{-0.05in%
}0, \\
\partial _{t}u^{c}+\partial _{x}p^{c}\hspace{-0.05in} &=&\hspace{-0.05in}0,
\\
fu^{c}+\partial _{y}p^{c}\hspace{-0.05in} &=&\hspace{-0.05in}0.
\end{eqnarray}%
Clearly, these waves propagate as nondispersive waves in the zonal
(east--west) direction---as if it were $f\equiv 0$---and are in geostrophic
balance between the Coriolis and pressure gradient forces in the meridional
(south--north) direction. From (\ref{v=0}a, b) it follows that
\end{subequations}
\begin{equation}
p^{c}=A(y)K(x-ct)\equiv cu^{c},
\end{equation}%
where $K(\cdot )$ is an arbitrary function. By virtue of (\ref{v=0}c) then
it follows $\mathrm{d}A/\mathrm{d}y+fA/c=0,$ whose solution is%
\begin{equation*}
A(y)\propto \mathrm{e}^{-\tint^{y}\mathrm{d}y\text{ }f(y)/c}=\mathrm{e}%
^{-(f_{0}y+\frac{1}{2}\beta y^{2})/c},
\end{equation*}%
which requires, except there where $f_{0}\equiv 0$ (i.e. the equator), the
presence of a zonal coast to be physically meaningful.

\begin{problem}
Consider the Kelvin waves in the so-called $f$ plane, i.e. with $\beta
\equiv 0.$
\end{problem}

\subsubsection{Vertical Structure}

Under the Boussinesq approximation the five fields of system (\ref{sis1})
remain independent, thereby removing no wave solutions. We can thus safely
consider the vertical structure equation (\ref{Boussinesq}), which we
rewrite in the form%
\begin{equation}
\mathrm{d}^{2}F/\mathrm{d}F^{2}+m^{2}(z)F=0  \label{F}
\end{equation}%
where%
\begin{equation}
m^{2}(z):=\left[ N^{2}(z)-\omega ^{2}\right] \left( c^{-2}-c_{\mathrm{s}%
}^{-2}\right) .  \label{m}
\end{equation}%
Equation (\ref{F}) can be understood in two different senses. Within the
realm of the WKB approximation, (\ref{m}) defines a local vertical
wavenumber, and a solution to (\ref{F}) oscillates like%
\begin{equation}
F(z)\sim \mathrm{e}^{\pm \mathrm{i}\tint^{z}\mathrm{d}z\,m(z)}.
\end{equation}%
The other sense is that of \textit{vertical normal modes}, in which (\ref{F}%
) is solved in the whole water column with boundary conditions
\begin{subequations}
\label{BC}
\begin{eqnarray}
F(-H)\hspace{-0.05in} &=&\hspace{-0.05in}0, \\
gF(0)\hspace{-0.05in} &=&\hspace{-0.05in}c^{2}\mathrm{d}F(0)/\mathrm{d}z.
\end{eqnarray}%
Condition (\ref{BC}a) comes from imposing $w^{\prime }=0$ at $z=-H$ where $%
H, $ which must be a constant, is the depth of the fluid in the reference
state. Condition (\ref{BC}b) comes from the fact that $p^{\prime }=g\zeta
^{\prime }$ at $z=0$, which means that the surface is isopycnic (i.e. the
density does not change on the surface). This way one is left with a classic
Sturm--Liouville problem. Making the incompressibility approximation and
assuming a uniform stratification in the reference state, namely $N=\bar{N}=%
\mathrm{const.},$ it follows that
\end{subequations}
\begin{equation}
\omega ^{2}=\bar{N}^{2}-mg\tan mH.
\end{equation}%
(Notice that to obtain $m$ is necessary to fix a value of $\omega .$) In the
hydrostatic limit, $\omega ^{2}\ll \bar{N}^{2},$ it follows that the
vertical normal modes result from
\begin{equation}
\tan mH=s/(mH),  \label{modos}
\end{equation}%
where $s:=\bar{N}^{2}H/g$, which is a measure of the stratification, is such
that $0<s<\infty $ by static stability. In the ocean $s$ is typically very
small, so for $s\ll 1$ from (\ref{modos}) it follows that
\begin{equation}
m_{i}=\left\{
\begin{array}{ll}
\bar{N}/\sqrt{gH} & \text{if }i=0, \\
i\pi /H & \text{if }i=1,2,\cdots .%
\end{array}%
\right.
\end{equation}%
The first mode is called the \textit{external }or \textit{barotropic }mode;
the rest of the modes are termed the \textit{internal }or \textit{baroclinic
}modes, which are well separated from the latter in what length scale
respects. More precisely, the \textit{Rossby radii of deformation} are
defined by $R_{i}:=\bar{N}/(m_{j}\left| f_{0}\right| );$ for the barotropic
mode $R_{0}=\sqrt{gH}/\left| f_{0}\right| $ whereas for the baroclinic modes
$R_{i}=\bar{N}H/(i\pi \left| f_{0}\right| )\equiv \sqrt{s}R_{0}/(i\pi )\ll
R_{0}.$ Finally, the \textit{rigid lid approximation }consists of making $%
w^{\prime }=0$ at $z=0,$ which formally corresponds to take the limit $%
g\rightarrow \infty $ in (\ref{BC}b). This approximation filters out the
barotropic mode since it leads to $\tan mH=0.$

\begin{problem}
Demonstrate that $p^{\prime }=g\zeta ^{\prime }$ at $z=0.$
\end{problem}

\subsubsection{General Dispersion Relation}

Upon eliminating $c$ between (\ref{l}) and (\ref{m}) it follows that
\begin{equation}
\fbox{$\dfrac{\mathbf{k}^{2}+\beta k/\omega }{\omega ^{2}-f^{2}}=\dfrac{m^{2}%
}{N^{2}-\omega ^{2}}+c_{\mathrm{s}}^{-2}$,}  \label{general}
\end{equation}%
which is a fifth-order polynomial in $\omega $ that constitutes the general
dispersion relation for linear ocean waves in an asymptotic WKB sense. This
is the main result of this paper. Approximate roots of (\ref{general}) are:
\begin{equation}
\text{acoustic}:\omega ^{2}=(\mathbf{k}^{2}+m^{2})c_{\mathrm{s}}^{2}\text{,}
\label{AW}
\end{equation}%
which holds for $\omega ^{2}\gg N^{2}$ (i.e. very high frequencies);

\begin{equation}
\text{Poincar\'{e}}:\omega ^{2}=\frac{\mathbf{k}^{2}N^{2}+m^{2}f^{2}}{%
\mathbf{k}^{2}+m^{2}},  \label{PW}
\end{equation}%
which follows upon taking the limit $c_{\mathrm{s}}^{-2}\rightarrow 0$ and
is valid for frequencies in the range $f^{2}<\omega ^{2}<N^{2}$; and%
\begin{equation}
\text{Rossby}:\omega =-\frac{\beta k}{\mathbf{k}^{2}+\left(
m^{2}/N^{2}\right) f^{2}},  \label{RW}
\end{equation}%
which also follows in the limit $c_{\mathrm{s}}^{-2}\rightarrow 0$ but is
valid for $\omega ^{2}\ll f^{2}$ (i.e. very low frequencies).

\begin{problem}
Demonstrate that the classical dispersion relations for Poincar\'{e} waves, $%
\omega ^{2}=f^{2}+c^{2}\mathbf{k}^{2}$, and surface gravity waves, $\omega
^{2}=g\left| \mathbf{k}\right| \tanh \left| \mathbf{k}\right| H,$ are
limiting cases of (\ref{PW}).
\end{problem}

\section{Discussion\label{discussion}}

The inviscid, unforced linearized equations of motion (\ref{sis1}) have
\textit{five} prognostic equations for \textit{five} independent dynamical
fields. As a consequence, \textit{five} is the number of waves sustained by (%
\ref{sis1}) which satisfy the general dispersion relation (\ref{general}) in
an asymptotic WKB sense. In proper limits, two acoustic waves (AW), two
Poincar\'{e} waves (PW), and one Rossby wave (RW) can be identified. The
fact that the number of waves supported by the linearized dynamics equals
the number of independent dynamical fields or prognostic equations, i.e. the
degrees of freedom of (\ref{sis1}), means that the waves constitute a
complete set of solutions of the linearized dynamics (cf. Table \ref{tabla}).

%TCIMACRO{
%\TeXButton{B}{\renewcommand{\arraystretch}{1}
%\begin{table}[t]
%\centering}}%
%BeginExpansion
\renewcommand{\arraystretch}{1}
\begin{table}[t]
\centering%
%EndExpansion
\begin{tabular}{c}
\hline
\\
$\left.
\begin{array}{c}
\partial _{t}\zeta ^{\prime }=\cdots \\
\partial _{t}p^{\prime }=\cdots \\
\partial _{t}\mathbf{u}^{\prime }=\cdots \\
\partial _{t}w^{\prime }=\cdots%
\end{array}%
\right\} :5\text{ independent fields }\leftrightarrow 5\text{ waves}:\left\{
\begin{array}{l}
\text{2 AW} \\
\text{2 PW} \\
\text{1 RW}%
\end{array}%
\right. $ \\
\\
$%
\begin{array}{cl}
\text{\textbf{incompressibility\thinspace }}\downarrow &
\begin{array}{l}
\partial _{z}w^{\prime }=-\mathbf{\nabla }\cdot \mathbf{u}^{\prime } \\
\Delta p^{\prime }=-\mathbf{\nabla }\cdot f\mathbf{\hat{z}}\times \mathbf{u}%
^{\prime }-\partial _{z}(N^{2}\zeta ^{\prime })%
\end{array}%
\end{array}%
$ \\
\\
$\left.
\begin{array}{c}
\partial _{t}\zeta ^{\prime }=\cdots \\
\partial _{t}\mathbf{u}^{\prime }=\cdots%
\end{array}%
\right\} :3\text{ independent fields }\leftrightarrow 3\text{ waves}:\left\{
\begin{array}{l}
\text{2 PW} \\
\text{1 RW}%
\end{array}%
\right. $ \\
\\
$%
\begin{array}{cc}
\text{\textbf{quasigeostrophy\thinspace }}\downarrow & \partial _{z}\mathbf{u%
}^{\prime }=\dfrac{N^{2}}{f_{0}}\mathbf{\hat{z}}\times \mathbf{\nabla }\zeta
^{\prime }%
\end{array}%
$ \\
\\
$\partial _{t}q^{\prime }=\cdots :1$ independent field $\leftrightarrow 1$
wave $:$ 1$\text{ RW}$ \\
\\ \hline
\end{tabular}%
\caption{Reduction of independent fields (and, hence, prognostic equations) by the
incompressibility and quasigeostrophic approximations. Here, AW, PW, and RW stand for acosutic waves,
Poincar\'e waves, and Rossby wave, respectively; $\Delta:=\nabla^2+
\partial_{zz}$ is the three-dimensional Laplacian; and
$q^{\prime }:=f+\nabla ^{2}p^{\prime }/f_{0}+\partial
_{z}(f_{0}N^{-2}\partial _{z}p^{\prime })$ is the so-called
quasigeostrophic potential vorticity.\label{tabla}}%
%TCIMACRO{
%\TeXButton{E}{\end{table}
%\renewcommand{\arraystretch}{1.5}}}%
%BeginExpansion
\end{table}
\renewcommand{\arraystretch}{1.5}%
%EndExpansion

The number of possible eigensolutions can be reduced is approximations that
eliminate some of the prognostic equations, or independent dynamical fields,
of the system are performed. The Boussinesq approximation, which is very
appropriate for the ocean, does not eliminate prognostic equations and has
the virtue of reducing the mathematical complexity of the governing
equations considerably. The incompressibility approximation, in turn,
removes two degrees of freedom: the vertical velocity is diagnosed by the
horizontal velocity,
\begin{equation}
\partial _{z}w^{\prime }=-\mathbf{\nabla }\cdot \mathbf{u}^{\prime }\mathbf{,%
}
\end{equation}%
and the latter along with the density diagnose the pressure field through
the three-dimensional Poisson equation
\begin{equation}
(\nabla ^{2}+\partial _{zz})p^{\prime }=-\mathbf{\nabla }\cdot f\mathbf{\hat{%
z}}\times \mathbf{u}^{\prime }-\partial _{z}(N^{2}\zeta ^{\prime }).
\end{equation}%
As a consequence, the two AW are filtered out and one is left with the two
PW and the RW. With these two approximations the Euler equations (\ref{sis1}%
) reduces to the so called \textit{Euler-Boussinesq equations}. Performing
in addition the hydrostatic approximation, which corresponds to neglecting $%
\partial _{t}w^{\prime }$ in Newton's vertical equation, does not amount to
a reduction of degrees of freedom because the vertical velocity is already
diagnosed by the horizontal velocity. In this case, the density field
diagnoses the pressure field through%
\begin{equation}
\partial _{z}p^{\prime }=-N^{2}\zeta ^{\prime }.
\end{equation}%
With this approximation (which implies that of incompressibility) and the
Boussinesq approximation, system (\ref{sis1}) reduces to what is known in
geophysical fluid dynamics as the \textit{primitive equations. }Finally, one
approximation that eliminates independent fields is the \textit{%
quasigeostrophic approximation}, which is often used to study low frequency
motions in the ocean, and the Earth and planetary atmospheres.\textit{\ }In
this approximation the density diagnoses the horizontal velocity through the
``thermal wind balance,''%
\begin{equation}
\partial _{z}\mathbf{u}^{\prime }=\dfrac{N^{2}}{f_{0}}\mathbf{\hat{z}}\times
\mathbf{\nabla }\zeta ^{\prime },
\end{equation}%
thereby removing two degrees of freedom and leaving only one RW.

\begin{acknowledgments}
The author has imparted lectures based on the present material to students
of the doctoral program in physical oceanography at CICESE (Ensenada, Baja
California, Mexico). Part of this material is inspired on a seminal homework
assigned by the late Professor Pedro Ripa. To his memory this article is
dedicated.
\end{acknowledgments}

\appendix

\section{Solutions to Some of the Problems\label{answers}}

%TCIMACRO{\TeXButton{theom_0}{\setcounter{theorem}{0}}}%
%BeginExpansion
\setcounter{theorem}{0}%
%EndExpansion

\begin{problem}
To describe the dynamics in a noninertial reference frame such as one tied
to the rotating Earth, two forces must be included: the Coriolis and
centrifugal forces. However, Laplace \cite{DeLaplace-1775} showed that if
the upward coordinate $z$ is chosen not to lie in the direction of the
gravitational attraction, but rather to be slightly tilted toward the
nearest pole, the centrifugal and gravitational forces can be made to
balance one another in a horizontal plane (cf. also Ref.
%TCIMACRO{\TeXButton{Ripa-FCE-97}{\onlinecite{Ripa-FCE-97}}}%
%BeginExpansion
\onlinecite{Ripa-FCE-97}%
%EndExpansion
). With this choice the Coriolis force is the only one needed to describe
the dynamics. Notice that the absence of the centrifugal force in a system
fixed to the Earth is what actually makes rotation effects real: they cannot
be removed by a change of coordinates.
\end{problem}

\begin{problem}
In the absence of diffusive processes, the isopycnal $z=\zeta $ is a
material surface, i.e. $[\mathbf{u}+\mathbf{\hat{z}\,}w-(\mathbf{u}_{\zeta }+%
\mathbf{\hat{z}\,}w_{\zeta })]\cdot \lbrack \mathbf{\nabla }\zeta -\mathbf{%
\hat{z}\,(}1-\partial _{z}\zeta )]=0.$ Here, $\mathbf{u}_{\zeta }+\mathbf{%
\hat{z}\,}w_{\zeta }$ denotes the velocity of \textit{some} point on the
surface [the velocity of a surface is not defined and it only makes sense to
speak of the velocity in a given direction, e.g. the normal direction, in
whose case it is $\mathbf{\hat{z}\,(}1-\partial _{z}\zeta )-\nabla \zeta $].
From the trivial relation $z-\zeta =0$ then it follows that $\left( \mathbf{u%
}_{\zeta }+\mathbf{\hat{z}\,}w_{\zeta }\right) \cdot \lbrack \mathbf{\nabla }%
\zeta -\mathbf{\hat{z}\,(}1-\partial _{z}\zeta )]=-\partial _{t}\zeta $ and,
hence, $\mathrm{D}\zeta =w$ at $z=\zeta .$
\end{problem}

\begin{problem}
To perform the linearization of the equations of motion, we write%
\begin{equation}
\begin{array}{ccccccc}
(\mathbf{u},w,\zeta ) & = &  &  & (\mathbf{u}^{\prime },w^{\prime },\zeta
^{\prime }) & + & \cdots , \\
(p,\rho ) & = & (\rho _{\mathrm{r}},p_{\mathrm{r}}) & + & (\rho ^{\prime
},p^{\prime }) & + & \cdots , \\
O & : & 1 &  & a &  & a^{2}%
\end{array}%
\end{equation}%
where $a$ is an infinitesimal amplitude. The $O(a)$ continuity equation (\ref%
{sis1}b) readily follows upon noticing that, up to $O(a),$ $c_{\mathrm{s}%
}^{-2}\mathrm{D}p=c_{\mathrm{s}}^{-2}(\partial _{t}p^{\prime }-\rho _{%
\mathrm{r}}gw^{\prime }).$ Up to $O(a),$ $\mathrm{D}\rho -c_{\mathrm{s}}^{-2}%
\mathrm{D}p=\partial _{t}\rho ^{\prime }-g^{-1}\rho _{\mathrm{r}%
}N^{2}w^{\prime }-c_{\mathrm{s}}^{-2}\partial _{t}p^{\prime }$ and $\mathrm{D%
}\zeta -w=\partial _{t}\zeta ^{\prime }-w^{\prime }.$ Then from the
relationships $\mathrm{D}\rho -c_{\mathrm{s}}^{-2}\mathrm{D}p=0$ and $%
\mathrm{D}\zeta -w=0$ it follows that $\rho ^{\prime }=c_{\mathrm{s}%
}^{-2}p^{\prime }+g^{-1}\rho _{\mathrm{r}}N^{2}\zeta ^{\prime }.$ Bearing in
mind the latter relation and the fact that $\mathrm{d}p_{\mathrm{r}}/\mathrm{%
d}z=-g\rho _{\mathrm{r}}$, the $O(a)$ vertical Newton's equation (\ref{sis1}%
d) then follows.%
%TCIMACRO{\TeXButton{theom_3}{\setcounter{theorem}{3}}}%
%BeginExpansion
\setcounter{theorem}{3}%
%EndExpansion
\end{problem}

\begin{problem}
For the ocean $c_{\mathrm{s}}\sim 1500$ $\mathrm{m}$ $\mathrm{s}^{-1}\gg
\sqrt{gH}\sim 200$ $\mathrm{m}$ $\mathrm{s}^{-1}$; by contrast, for the
atmosphere $c_{\mathrm{s}}\sim 350$ $\mathrm{m}$ $\mathrm{s}^{-1}\sim \sqrt{%
gH}$ with $H\sim 12$ \textrm{km}, which is the typical height of the
troposphere.%
%TCIMACRO{\TeXButton{theom_6}{\setcounter{theorem}{6}}}%
%BeginExpansion
\setcounter{theorem}{6}%
%EndExpansion
\end{problem}

\begin{problem}
At the surface $z=\eta $ it is $w=\partial _{t}\eta +\mathbf{u}\cdot \mathbf{%
\nabla }\eta $ and $p=0$ (here, $p$ is a kinematic pressure, i.e. divided by
a constant reference density $\bar{\rho}$). Writing $\eta =\eta ^{\prime
}+O(a^{2})$ and Taylor expanding about $z=0$ it follows, on one hand,
\begin{equation}
w^{\prime }+\eta ^{\prime }\partial _{z}w^{\prime }+O(a^{3})=\partial
_{t}\eta ^{\prime }+\mathbf{u}^{\prime }\cdot \mathbf{\nabla }\eta ^{\prime
}+O(a^{3})  \label{w}
\end{equation}%
at $z=0$, and, on the other hand,%
\begin{equation}
p_{\mathrm{r}}+(p_{\mathrm{r}}+\mathrm{d}p_{\mathrm{r}}/\mathrm{d}z)\eta
^{\prime }+\eta ^{\prime }\partial _{z}p^{\prime }+O(a^{3})=0  \label{p}
\end{equation}%
at $z=0.$ From (\ref{w}) it follows, to the lowest order, $w^{\prime
}=\partial _{t}\eta ^{\prime }$ at $z=0$. Since $w^{\prime }=\partial
_{t}\zeta ^{\prime },$ for a wave (i.e. $\partial _{t}\neq 0$) then it
follows that $\eta ^{\prime }=\zeta ^{\prime }$ at $z=0.$ Taking into
account the latter and choosing $\bar{\rho}=\rho _{\mathrm{r}}(0)$, from (%
\ref{p}) it follows, to the lowest order, $p^{\prime }=g\eta ^{\prime
}\equiv g\zeta ^{\prime }$ at $z=0$ since $p_{\mathrm{r}}=0$ and \textrm{d}$%
p_{\mathrm{r}}/\mathrm{d}z=-g\rho _{\mathrm{r}}/\bar{\rho}\equiv -g$ at $%
z=0. $
\end{problem}

\begin{problem}
The classical dispersion relation for Poincar\'{e} waves corresponds to the
hydrostatic limit, which requires $m^{2}\gg \mathbf{k}^{2}$ (i.e. that the
vertical length scales be shorter than the horizontal length scales). Under
this conditions, $\omega ^{2}=f^{2}+\mathbf{k}^{2}N^{2}/m^{2}=f^{2}+c^{2}%
\mathbf{k}^{2}$. To obtain the dispersion relation for surface gravity waves
one needs to take into account boundary conditions (\ref{BC}): making $%
N^{2}\equiv 0\equiv f^{2}$ it follows, on one hand, that $m^{2}=-\mathbf{k}%
^{2},$ and, on other, that $\omega ^{2}=-mg\tan mH.$ The dispersion relation
$\omega ^{2}=g\left| \mathbf{k}\right| \tanh \left| \mathbf{k}\right| H$
then readily follows.
\end{problem}

\bibliographystyle{apsr}

\end{document}